\documentclass{elsart}
\usepackage{amssymb,epsfig,multicol}

\def\fq{{f^q}}
\def\pq{{p^q}}
\def\dm{{\partial_\mu}}
\def\be{{\begin{equation}}}
\def\ee{{\end{equation}}}

\begin{document}
\begin{frontmatter}

\title{Relativistic Nonextensive Thermodynamics}
\author{A. Lavagno}
\address{
Dipartimento di Fisica, Politecnico di Torino and INFN, Sezione di Torino}
\address{C.so Duca degli Abruzzi 24, I-10129 Torino, Italy}
\date{\today}
\maketitle

\begin{abstract}
Starting from the basic prescriptions of the Tsallis' nonextensive
thermostatistics, i.e. generalized entropy and normalized
$q$-expectation values, we study the relativistic nonextensive
thermody\-na\-mics and derive a Boltzmann transport equation that
implies the validity of the $H$-theorem where a local nonextensive
four-entropy density is considered. Macroscopic thermodynamic
functions and the equation of state for a perfect gas are
derived at the equilibrium.

\noindent
{\it PACS:} 05.20.Dd, 05.70.Ln, 05.90.+m, 25.75.-q \\
%\noindent
%{\it Keywords:} Relativistic thermostatistics, Nonextensive statistics}

\end{abstract}
\end{frontmatter}

%\begin{multicols}{2}

\section{Introduction}

Recently, there is an increasing evidence that the generalized
nonextensive statistical mechanics, proposed by Tsallis
\cite{tsallis}, can be considered as the more appropriate basis of a
theoretical framework to deal with physical phenomena where
long-range interactions, long-range microscopic memories and/or
fractal space-time constraints are present (cf. \cite{tsallis} for
details). A considerable variety of physical applications involve
a quantitative agreement between experimental data and theoretical
models based on Tsallis' thermostatistics \cite{biblio}. In
particular there is a growing interest to high energy physics
applications of nonextensive statistics. Several authors outline
the possibility that experimental observations in relativistic
heavy-ion collisions can reflect nonextensive features during the
early stage of the collisions and the thermalization evolution of
the system \cite{albe,wilk,rafe,bediaga,beck}.

The basic aim of this letter is to study the nonextensive
statistical mechanics formalism in the relativistic regime and to 
investigate, through an appropriate relativistic Boltzmann
equation, the non-equilibrium and the equilibrium thermodynamics
relations.

\section{Basic assumptions in nonextensive thermostatistics}

Let us briefly review some basic assumptions of the 
nonextensive thermostatistics that will be useful in view of
the relativistic extension.

Starting point of the Tsallis' generalization of the
Boltzmann-Gibbs statistical mechanics is the introduction of a
$q$-deformed entropy functional defined, in a phase space system,
as \cite{tsallis}
\begin{equation}
S_q=-k_{_B} \int d\Omega \, \pq \ln_q p \; , \label{entropy}
\end{equation}
where $k_{_B}$ is the Boltzmann constant, $p=p(x,v)$ is the phase
space probability distribution, $d\Omega$ stands for the
corresponding phase space volume element and $\ln_q
x=(x^{1-q}-1)/(1-q)$ is, for $x>0$, the $q$-deformed logarithmic
function. For the real parameter $q\rightarrow 1$,
Eq.(\ref{entropy}) reduces to the standard Boltzmann-Gibbs entropy
functional.

In the equilibrium canonical ensemble, under the constraints
imposed by the probability normalization
\begin{equation}
\int d\Omega \, p=1 \; ,
\end{equation}
and the normalized $q$-mean expectation value of the energy
\cite{mendes}
\begin{equation}
\langle E \rangle_q=\frac{\int d\Omega \,\pq H(x,v)}{\int
d\Omega\, \pq} \; ,
\end{equation}
the maximum entropy principle gives the probability distribution
\cite{mendes,pla1}
\begin{equation}
p(x,v)=\frac{f(x,v)}{Z_q}\; , 
\label{proba}
\end{equation}
where
\begin{equation}
f(x,v)=[1-(1-q)\beta (H(x,v)-\langle H \rangle_q)]^{1/(1-q)}  \; 
\label{distri}
\end{equation}
and 
\begin{equation}
Z_q=\int d\Omega\, f(x,v) \; .
\label{parti}
\end{equation}

Let us note that, depending from the extremization procedure, in
Ref.\cite{mendes} the above factor $\beta$ is only proportional to
the Lagrange multiplier, because of the probability distribution
is self-referential, while, in Ref.\cite{pla1}, $\beta$ is
actually the Lagrange multiplier associated to the energy
constraint.

By using the definitions in Eqs. (\ref{proba}) and (\ref{parti})
into the relation $Z_q^{1-q}=\int d\Omega\;\pq$ \cite{pla1}, 
the following identity holds \cite{pla2}
\begin{equation}
\int d\Omega \, f(x,v)\equiv \int d\Omega \,\fq(x,v)  \; ,
\label{rela}
\end{equation}
and the normalized $q$-mean
expectation value for a physical observable $A(x,v)$ can be
expressed as
\begin{equation}
\langle A \rangle_q=\frac{\int d\Omega \,\fq\, A(x,v)}{\int
d\Omega\, \fq}\equiv \frac{1}{Z_q} \int d\Omega \,\fq\, A(x,v)\; .
\label{mean}
\end{equation}
Therefore, the probability distribution and the $q$-mean value of an
observable have the same normalization factor $Z_q$, as in the
extensive statistical mechanics. 
Such a non-trivial property does not depend
on the equilibrium frame but it comes
from the normalization condition and holds at any time (if we
require that the transport equation conserves the probability
normalization or the number of particles). This observation will
play a crucial role in the correct formulation of the relativistic
Boltzmann equation and in the definition of the thermodynamic variables.

\section{Relativistic kinetic theory}

On the basis of the above prescriptions, we are able to progress in the 
formulation of the relativistic nonextensive statistical mechanics. 
Let us start defining the basic macroscopic variables in the language
of relativistic kinetic theory.
 Because we are going to describe a non-uniform system in the
phase space, we introduce the particle four-flow as
\begin{equation}
N^\mu(x)=\frac{1}{Z_q}\int \frac{d^3p}{p^0} \, p^\mu \,f(x,p) \; ,
\label{nmu}
\end{equation}
and the energy-momentum flow as
\begin{equation}
T^{\mu\nu}(x)=\frac{1}{Z_q}\int \frac{d^3p}{p^0} \, p^\mu p^\nu \,
f^q(x,p) \; , \label{tmunu}
\end{equation}
where we have set $\hbar=c=1$, $x\equiv x^\mu=(t,{\bf x})$,
$p\equiv p^\mu=(p^0,{\bf p})$ and $p^0=\sqrt{{\bf p}^2+m^2}$ is
the relativistic energy. The four-vector $N^\mu=(n,{\bf
j})$ contains the probability density $n=n(x)$ (which is
normalized to unity) and the probability flow ${\bf j}={\bf
j}(x)$. The energy-momentum tensor contains the normalized
$q$-mean expectation value of the energy density, as well as the energy flow,
the momentum and the momentum flow per particle. Its expression follows
directly from the definition (\ref{mean}); for this reason it is
given in terms of $f^q(x,p)$.

In order to derive a relativistic Boltzmann equation for a
dilute system in nonextensive statistical mechanics, we consider
the finite volume elements $\Delta^3x$ and $\Delta^3p$ in the
phase space. These volume elements are large enough to contain a
very large number of particles but also small enough compared to the
macroscopic dimension of the system. If, in the volume $\Delta^3x
\Delta^3p$, is contained a representative sample of system, we can
assume that Eq.(\ref{rela}) still holds in such phase space
portion. Then, in the Lorentz frame, the particle fraction $\Delta
N(x,p)$ in the volume $\Delta^3x \Delta^3p$ can be written
as
\begin{equation}
\Delta N(x,p)=\frac{N}{Z_q} \int_{\Delta^3x}
\int_{\Delta^3p} d\Omega \, f^q(x,p) \; ,
\end{equation}
where $N$ is the total number of particles of the system.
With respect to the observer's frame of reference, the above
expression becomes
\begin{equation}
\Delta N(x,p)=\frac{N}{Z_q} \int_{\Delta^3\sigma}
\int_{\Delta^3p} d^3\sigma_\mu \frac{d^3p}{p^0} p^\mu f^q(x,p)\; ,
\end{equation}
where $d^3\sigma_\mu$ is a time-like three-surface element of a
plane space-like surface $\sigma$ \cite{groot}. Requiring that the net
flow through the surface $\Delta^3\sigma$ of the element $\Delta^4
x$ vanishes in absence of collisions, we have: $p^\mu
\dm\fq(x,p)=0$. While considering collisions between particles,
the Boltzmann equation becomes
\begin{equation}
p^\mu \dm\fq(x,p)=C_q(x,p)  \; , \label{boltz}
\end{equation}
where $C_q(x,p)$ is the $q$-deformed collision term that, under
the hypothesis that only binary collisions occur in the gas, 
can be expressed as
\begin{eqnarray}
C_q(x,p)=&&\frac{1}{2}\! \int\!\!\frac{d^3p_1}{p^0_1}
\frac{d^3p{'}}{p{'}^0} \frac{d^3p{'}_1}{p{'}^0_1} \Big
\{h_q[f{'},f{'}_1]  W(p{'},p{'}_1\vert p,p_1) \nonumber \\
&&-h_q[f,f_1]  W(p,p_1\vert p{'},p{'}_1) \Big\}\; .
\end{eqnarray}
In the above equation, we have set $W(p,p_1\vert p{'},p{'}_1)$ as the
transition rate between two particle state with initial four-momentum
$p$ and $p{'}$ and a final state with four-momenta $p_1$ and $p{'}_1$;
$h_q[f,f_1]$ is the correlation function related to two particles
in the same space-time position but with different four-momenta
$p$ and $p_1$, respectively. The factorization of $h_q$ in two
single probability distributions (uncorrelated particles at the
same spatial point) is the celebrated hypothesis of molecular
chaos (Boltzmann's Stosszahlansatz). Thus, the function $h_q$
defines implicitly a generalized nonextensive molecular chaos
hypothesis.

By assuming the conservation of the energy-mo\-men\-tum in the
collisions (i.e. $p^\mu+p{'}^\mu=p^\mu_1+p{'}^\mu_1$) and requiring that
the correlation function $h_q$ is symmetric and always positive 
($h_q[f,f_1]=h_q[f_1,f]$, $h_q[f,f_1]>0$), it is
easy to show that collision term satisfies the following property
\begin{equation}
{F}[\psi]=\int \frac{d^3p}{p^0}\, \psi(x,p)\, C_q(x,p)=0 \; ,
\label{law}
\end{equation}
if
\begin{equation}
\psi(x,p)=a(x)+b_\mu(x)p^\mu  \; , \label{abp}
\end{equation}
where $a(x)$ and $b(x)$ are arbitrary functions.

By choosing $\psi=const.$ and using the Boltzmann equation (\ref{boltz}),
then Eq.(\ref{law}) implies
\begin{equation}
\frac{\partial}{\partial t} \int d\Omega \fq(x,p) =0\; ,
\end{equation}
and this is, on account of Eq.(\ref{rela}), nothing else that the
conservation of the probability normalization $Z_q$.
%\begin{equation}
%N^{*\mu}(x)=\frac{1}{Z_q} \int\frac{d^3p}{p^0} p^\mu \fq  \; .
%\end{equation}
Otherwise, by setting $\psi=b_\mu p^\mu$, we have from Eq.(\ref{law})
\begin{equation}
\partial_\nu T^{\mu\nu}(x)=0  \; ,
\end{equation}
which implies the energy and the momentum conservation.

Let us remark that to have conservation of the
probability normalization, energy and momentum, it is
crucial that not only the collision term $C_q$ be explicitly
deformed by means of the function $h_q$,
but also the streaming term $p^\mu \dm\fq$. This matter of fact
is a directly consequence of the nonextensive statistical
prescription of the normalized $q$-mean expectation value
(\ref{mean}) and is not taken into account in the non-relativistic 
formulation of Ref.\cite{lima}.

\section{Local $H$-theorem}

The relativistic local $H$-theorem states that the entropy
production $\sigma_q (x)=\dm S_q^\mu(x)$ at any space-time point
is never negative.

Assuming the validity of the Tsallis entropy (\ref{entropy}), it
appears natural to introduce the nonextensive four-flow entropy 
$S_q^\mu(x)$ as follows
\begin{equation}
S_q^\mu(x)=-k_{_B} \,\int \frac{d^3p}{p^0} \,p^\mu f^q(x,p) [\ln_q
f(x,p)-1] \; . \label{entro4}
\end{equation}
On the basis of the above equation the entropy production can be written as
\begin{equation}
\sigma_q(x)=-k_{_B}  \int\frac{d^3p}{p^0} \ln_q f\, \, p^\mu
\dm\fq \equiv -k_{_B} F[\ln_q f] \; , \label{sigma}
\end{equation}
where the second identity follows from the
Boltzmann equation (\ref{boltz}) and the definition of 
$F[\psi]$ in Eq.(\ref{law}). After simple manipulations,
Eq.(\ref{sigma}) can be rewritten as
\begin{eqnarray}
\sigma_q(x)=&&\frac{k_{_B}}{8} \int\frac{d^3p_1}{p^0_1}
\frac{d^3p_1}{p^0_1} \frac{d^3p{'}}{p{'}^0}
\frac{d^3p_1{'}}{p{'}^0_1} \Big( \ln_q f{'}+\ln_q f{'}_1- 
\ln_q f+\ln_q f_1\Big) \times \nonumber \\
&&\Big \{h_q[f{'},f{'}_1]\, W(p{'},p{'}_1\vert p,p_1)- 
h_q[f,f_1] \, W(p,p_1\vert p{'},p{'}_1) \Big\}\; .
\end{eqnarray}

By assuming the detailed-balance property $W(p,p_1\vert
p{'},p{'}_1)=W(p{'},p{'}_1\vert p,p_1)$, we have that the entropy
production is always an increasing function,
if $q>0$ and if the function $h_q$ satisfies the general condition
\begin{equation}
h_q[f,f_1]=h_q[\ln_q f+\ln_q f_1] \; ,
\end{equation}
in addition to be symmetric and always positive.

Because the nonextensive formalism reduces to the standard
Boltzmann kinetic formulation for $q \rightarrow 1$, it appears
natural to postulate the $q$-generalized Boltzmann molecular chaos
hypothesis as
\begin{equation}
h_q[f,f_1]=e_q(\ln_q f+\ln_q f_1)  \; , \label{ansatz}
\end{equation}
where we have introduced the Tsallis $q$-exponential function
\begin{equation}
e_q(x)=[1+(1-q)x]^{1/(1-q)} \; , 
\end{equation}
which satisfies the properties:
$e_q(\ln_q x)=x$ and $e_q(x)\cdot e_q(y)=e_q[x+y+(1-q)xy]$. Let us
note that a similar expression for the function $h_q$ was
previously introduced in Ref.s\cite{lima,kania} and a rigorous
justification of the validity of the ansatz (\ref{ansatz}) can be
found only by means of a microscopic analysis of the dynamics of
correlations in nonextensive statistics.

\section{Equilibrium and equation of state}

The condition that entropy production vanishes eve\-ry\-where,
together with requirement that the equilibrium probability
distribution $f_{eq}$ must be a solution of the transport equation
(\ref{boltz}), uniquely defines the state of equilibrium. Taking
into account of Eqs.(\ref{law}), (\ref{abp}) and (\ref{sigma}),
the condition $\sigma_q=0$ can only occur when $\ln_q
f_{eq}=a+b_\mu p^\mu$. By imposing that $f_{eq}$ must satisfy
Eq.(\ref{boltz}) and after simple redefinition of the
coefficients $a$ and $b$, the equilibrium probability distribution
can be written as a Tsallis-like distribution
\begin{equation}
f_{eq}(p)= \frac{1}{Z_q}\left [1-(1-q) \frac{p^\mu U_\mu}{k_{_B}T}
\right]^{1/(1-q)} \; ,
\end{equation}
where $U_\mu$ is the hydrodynamic four-velocity \cite{groot} and
$f_{eq}$ depends only on the momentum in absence of an external
field. At this stage, $k_{_B}T$ is a free parameter and only in
the derivation of the equation of state it will be identified with
the physical temperature. Moreover, it is easy to show that
$f_{eq}$ is a solution of the transport equation (\ref{boltz}).

We are able now to evaluate explicitly all other thermodynamic
variables and provide a complete macroscopic description of a
relativistic system at the equilibrium. Let us first calculate the
probability density defined as
\begin{equation}
n=N^\mu U_\mu=\frac{1}{Z_q}\int\frac{d^3p}{p^0} \, p^\mu U_\mu\,
f_{eq}(p)\;  .
\end{equation}
Since $n$ is a scalar, it can be evaluated in the rest frame where
$U^\mu=(1,0,0,0)$. Setting $\tau=p^0/k_{_B}T$ and $z=m/k_{_B}T$,
the above integral can be written as
\begin{eqnarray}
%\;\;\;\;\;\;\;\;\;\;\;\;\;\;\;\;\;\;\;\;\;\;\;\;
n&=&\frac{4\pi}{Z_q} \, (k_{_B}T)^3 \int^\infty_z \!\!d\tau
(\tau^2-z^2)^{1/2} \, \tau \, e^{-\tau}_q \nonumber \\
%\;\;\;\;\;\;\;\;\;\;\;\;\;\;\;\;\;\;\;\;\;\;\;\;
&=&\frac{4\pi}{3\,Z_q} \, (k_{_B}T)^3 \int^\infty_z
\!\!d\tau (\tau^2-z^2)^{3/2} \, \left(e^{-\tau}_q\right)^q \; ,
\end{eqnarray}
where the last identity has been obtained by a partial
integration. Let us introduce the $q$-modified Bessel function of the second 
kind as follows
\begin{equation}
K_n(q,z)=\frac{2^n n!}{(2n)!}\frac{1}{z^n}\int^\infty_z
\!\!d\tau (\tau^2-z^2)^{n-1/2}\,
\left(e^{-\tau}_q\right)^q \; , \label{qbessel}
\end{equation}
then, the particle density can be cast into the compact form
\begin{equation}
n= \frac{4 \pi}{Z_q} \, m^2 \, k_{_B}T \, K_2(q,z) \;
.\label{densi}
\end{equation}

Similarly we can obtain the other macroscopic thermodynamic
variables. Considering the decomposition of the energy-momentum
tensor \cite{groot}: $T^{\mu\nu}=\epsilon\, U^\mu U^\nu-p\,
\Delta^{\mu\nu}$, where $\epsilon$ is the energy density, $p$ the
pressure and $\Delta^{\mu\nu}=g^{\mu\nu}-U^\mu U^\nu$, the
equilibrium pressure can be calculated as
\begin{equation}
p=-\frac{1}{3} T^{\mu\nu}\Delta_{\mu\nu}=-\frac{1}{3\,Z_q}
\int\frac{d^3p}{p^0} p^\mu p^\nu
\Delta_{\mu\nu}\fq_{\!\!\!\!\!eq}(p) \; ,
\end{equation}
and can be expressed as
\begin{equation}
p=\frac{4\pi}{Z_q}\,m^2\,(k_{_B}T)^2\, K_2(q,z) \; . \label{press}
\end{equation}
Comparing Eq.(\ref{densi}) with Eq.(\ref{press}), we obtain 
\begin{equation}
p=n\,k_{_B}T  \; ,
\end{equation}
which is the equation of state of a perfect gas if we identify $T$
as the physical temperature of the system. 
An explicit $q$-dependence does not appear. The same equation has
been derived in non-relativistic regime and in nonextensive
scenario \cite{pla2,abe}. 

We proceed now to calculate the energy density $\epsilon$ as 
\begin{equation}
\epsilon=T^{\mu\nu}U_\mu U_\nu=\frac{1}{Z_q}  \int\frac{d^3p}{p^0}
(p^\mu U_\mu)^2 \fq_{\!\!\!\!\!eq}(p) \; .
\end{equation}
Inserting the previously defined variables $\tau$ and $z$ and
using the definition in Eq.(\ref{qbessel}), we obtain
\begin{equation}
\epsilon=\frac{4\pi}{Z_q}\, m^4 \left [
3\frac{K_2(q,z)}{z^2}+\frac{K_1(q,z)}{z}\right ] \; .
\end{equation}
Thus the energy per particle $e=\epsilon/n$ is
\begin{equation}
e=3 \,k_{_B}T +m \,\frac{K_1(q,z)}{K_2(q,z)} \; ,
\end{equation}
which has the same structure of the relativistic expression
obtained in the framework of the equilibrium Boltzmann-Gibbs
statistics \cite{groot}.

In the non-relativistic limit ($p\ll 1$) the energy per particle
reduces to the well-known expression
\begin{equation}
e\simeq m+\frac{3}{2}\,k_{_B}T  \; .
\end{equation}
Also in this case, no explicit $q$-dependence is detected.

\section{Conclusion}

The physical motivation of this investigation lies in the strong
relevance that nonextensive statistics could have in high energy
physics. In this letter we have studied the thermostatistics of a 
relativistic system in nonextensive statistics; 
the obtained results can be easily extended to the 
case where an external force is present.

The prescription of the normalized $q$-mean expectation values
implies a consistent nonextensive generalization of the
macroscopic variables $N^\mu$ and $T^{\mu\nu}$. On this basis, we
have derived a generalized Boltzmann equation where both the streaming
and the collision terms depend on the deformation parameter
$q$. Such a transport equation conserves the probability
normalization (or number of particles) and is consistent with the
energy-momentum conservation laws. The collision term contains a
generalized expression of the molecular chaos and for $q>0$
implies the validity of a generalized $H$-theorem, if the
nonextensive local four-density entropy (\ref{entro4}) is assumed.
At the equilibrium, the solution of the Boltzmann equation is a
relativistic Tsallis-like distribution and the equation of state
of a classical relativistic gas in nonextensive statistical
mechanics has the same form as in ordinary Boltzmann-Gibbs frame.

Finally, nonextensive statistical effects
can be detected in connection to
microscopic observables such as particle distribution, correlation
functions, fluctuations of thermodynamical variables but not directly 
in connection to macroscopic variables, such
as temperature or pressure, because the equation of state and the
macroscopic thermodynamical relations do not explicitly depend on the
deformation parameter $q$.\\

\noindent
{\bf Acknowledgements}\\

It is a pleasure to thank P. Quarati and C. Tsallis for useful discussions.

%\end{multicols}


\begin{thebibliography}{30}

\bibitem{tsallis}
C. Tsallis, J. Stat. Phys. {\bf 52}, 479 (1988). See also: {\it
Nonextensive Statistical Mechanics and Its Applications}, Editors
S. Abe, Y. Okamoto, Springer Verlag, 2001.
\bibitem{biblio}
See http://tsallis.cat.cbpf.br/biblio.htm for a regularly updated
bibliography on the subject.
\bibitem{albe}
W.M. Alberico, A. Lavagno, P. Quarati, Eur. Phys. J. C 12 (2000)
499; W.M. Alberico, A. Lavagno, P. Quarati, Nucl. Phys. A 680 (2001)
94; A. Lavagno, P. Quarati, Chaos, Solitons and Fractals 13 (2002)
569; A. Lavagno, Physica A 305 (2002) 238.
\bibitem{wilk}
G. Wilk, Z. Wlodarczyk, Phys. Rev. Lett. 84 (2000) 2770; O.V.
Utyuzh, G. Wilk, Z. Wlodarczyk, J. Phys. G 26 (2000) L39.
\bibitem{rafe}
D.B. Walton, J. Rafelski, Phys. Rev. Lett. 84 (2000) 31.
\bibitem{bediaga}
I. Bediaga, E.M.F. Curado, J.M. de Miranda, Physica A 286 (2000)
156.
\bibitem{beck}
C. Beck, Physica A 286 (2000) 164.
\bibitem{mendes}
C. Tsallis, R.S. Mendes, A.R. Plastino, Physica A 261 (1998) 534.
\bibitem{pla1}
S. Mart\'inez, F. Nicol\'as, F. Pennini, A. Plastino, Physica A
286 (2000) 489.
\bibitem{pla2}
S. Mart\'inez, F. Pennini, A. Plastino, Phys. Lett. A 278 (2000)
47.
\bibitem{groot}
S.R. Groot, W.A. van Leeuwen, Ch. G. van Weert, {\it
Re\-la\-ti\-vi\-stic kinetic theory}, North-Holland, 1980.
%B.M. Boghosian, Bras. J. Phys. (1999) ???????
\bibitem{lima}
J.A.S. Lima, R. Silva, A.R. Plastino, Phys. Rev. Lett. 86 (2001)
2938.
\bibitem{kania}
G. Kaniadakis, Physica A 296 (2001) 405; Phys. Lett. A 288 (2001) 283.
\bibitem{abe}
S. Abe, S. Mart\'inez, F. Pennini, A. Plastino, Phys. Lett. A 281
(2001) 126.

\end{thebibliography}
\end{document}